\documentclass[12pt]{article}
\usepackage[utf8]{inputenc}
\pdfoutput=1
\usepackage{amsmath}
\usepackage{amssymb}
\usepackage{amsthm}

\usepackage[T1]{fontenc}
\usepackage{mathptmx}
\usepackage[default,scale=0.95]{lato}

\usepackage[left=1in,right=1in,top=1in,bottom=1in]{geometry}
\usepackage[singlespacing]{setspace}
\usepackage{parskip}
\usepackage[inline]{enumitem}
\setlist{nosep}
\providecommand{\keywords}[1]{\textbf{\textit{Keywords}:} #1}
\providecommand{\jel}[1]{\textbf{\textit{JEL Classification}:} #1}

\usepackage[dvipsnames]{xcolor}
\usepackage{graphics}
\usepackage[outdir=./]{epstopdf}
\usepackage{booktabs}
\usepackage{mathtools}
\mathtoolsset{multlined-width=0.9\linewidth,multlined-pos=b}
\usepackage{authblk}
\usepackage{tabularx}
\usepackage[justification=centering]{caption}

\usepackage[colorlinks=true,citecolor=blue,linkcolor=blue]{hyperref}

\usepackage[american]{babel}
\usepackage{csquotes}
\usepackage[style=apa]{biblatex}
\newcounter{theorem}
\numberwithin{theorem}{section}
\numberwithin{equation}{section}

\theoremstyle{definition}

\theoremstyle{plain}
\newtheorem{thm}[theorem]{Theorem}

\theoremstyle{remark}

\newcommand{\E}{\mathbb{E}} % Expectation
\newcommand{\e}{\mathrm{e}} % Exponential function

\begin{document}

\title{\bf Correlation Estimation in Hybrid Systems}
\author{Baron Law}
\affil{Agam Capital Management, LLC}
\date{July 6, 2023}
\maketitle	

\begin{abstract}
A simple method is proposed to estimate the instantaneous correlations between state variables in a hybrid system from the empirical correlations between observable market quantities such as spot rates, stock prices and implied volatilities. The new algorithm is extremely fast since only low-dimension linear systems are involved. If the resulting matrix from the linear systems is not positive semidefinite, the shrinking method,  which requires only bisection-style iterations, is recommended to convert the matrix to positive semidefinite. The square of short-term at-the-money implied volatility is suggested as the proxy for the unobservable stochastic variance. When the implied volatility is not available, a simple trick is provided to fill in the missing correlations. Numerical study shows that the estimates are reasonably accurate, when using more than 1,000 data points. In addition, the algorithm is robust to misspecified interest rate model parameters and the short-sampling-period assumption. G2++ and Heston are used for illustration, but the method can be extended to affine term-structure, local volatility and jump-diffusion models, with or without stochastic interest rate.
\end{abstract}
\jel{C13, G13}\\

\section{Introduction}

A hybrid system consists of multiple correlated stochastic models on interest rates, equities and foreign exchanges (FX). There are decent number of papers devoted to the pricing of derivatives under a hybrid model \parencite{34,35,36}, but very few of them discuss calibration issues, such as the problem of estimating correlations across multiple asset classes.

First of all, benchmark instruments are usually not very sensitive to cross correlations. For example, the price of a S\&P index option does not depend on the correlation between S\&P and EURUSD exchange rate, while its sensitivity on correlation between S\&P and USD interest rate is usually quite low. There are few OTC correlation-sensitive exotic options, but they are not liquidly traded.

Therefore, it is common to estimate the cross correlations from the observed time series \parencite{7,13}. The instantaneous correlation matrix is chosen such that the differences between the theoretical and empirical correlations of some output variables are minimized. Often, the theoretical correlation matrix exploits some well-known parametrization \parencite{21} to ensue positive semidefiniteness.

However, the diagonal blocks of the correlation matrix are usually calibrated from the volatility surfaces, so equality constraints are required in order to preserve the inner correlations. Optimization with equality constraints are often problematic; sometimes, the optimizer cannot even find a feasible candidate, let alone the optimum.

In this article, the approach in \parencite{7}, which deals only with equity, will be extended to multiple asset classes, represented by G2++ \parencite{14} and Heston \parencite{15}. Furthermore, the dilemma of unobservable stochastic variance and the positive semidefiniteness of the empirical matrix will also be addressed.

This paper is organized as follows: after the representative models of G2++ and Heston are defined in Section \ref{sec:comp}, the estimation algorithm will be presented in detail in Section \ref{sec:est}. The topic of unobservable stochastic variance will be discussed in Section \ref{sec:miss}. In Section \ref{sec:trans}, a method called shrinking will be exploited to convert the empirical matrix to positive semidefinite. Simple extensions to affine term-structure, local volatility, jump-diffusion, and stochastic interest rate models will be described in Section \ref{sec:ext}. Section \ref{sec:num} provides some insights to the accuracy and robustness of the method by comparing the estimated correlations with known values, while Section \ref{sec:con} concludes.

\section{Component Models} \label{sec:comp}
The new calibration approach will be illustrated using G2++ \parencite{14} and Heston \parencite{15}, but it will be extended to affine term structure \parencite{16}, local volatility \parencite{37} and Bates \parencite{18} in the later section.

Technically speaking, all the models should be in the same probability measure for the instantaneous correlations between Brownian motions to be well-defined; however, one knows from Girsanov theorem \parencite{20} that an absolutely continuous change of measure only affects the drift of a process, so the choice of measure for each model is irrelevant.

To be more precise, suppose $W_1^{Q1}(t)$ and $W_2^{Q2}(t)$ are Brownian motions under the measure $Q1$ and $Q2$ respectively. By Girsanov theorem, $dW_1^{Q2}(t) = dW_1^{Q1}(t) + X(t) dt$ is a Brownian motion under $Q2$, where $X(t)$ is the corresponding Girsanov kernel. The compensator process between $dW_1^{Q2}$ and $dW_2^{Q2}$ is given by
\begin{align}
	\left< dW_1^{Q2}, \ dW_2^{Q2} \right>_t
	&= \left< dW_1^{Q1} + X(\bullet) d\bullet, \ dW_2^{Q2} \right>_t \notag \\
	&= \left< dW_1^{Q1}, \ dW_2^{Q2} \right>_t  + \left< X(\bullet) d\bullet, \ dW_2^{Q2} \right>_t  \label{fv} \notag \\
	&= \left< dW_1^{Q1}, \ dW_2^{Q2} \right>_t
\end{align}
The second term in equation \eqref{fv} vanishes as $X(t) dt$ has finite variation. Hence,
\begin{align}
	\text{Corr}\left(dW_1^{Q2}(t), dW_2^{Q2}(t)\right) 
	&= \E\left(\left<dW_1^{Q2},dW_2^{Q2}\right>_t\right) \Big/ dt \notag \\
	&= \E\left(\left<dW_1^{Q1},dW_2^{Q2}\right>_t\right) \Big/ dt \notag \\
	&= \text{Corr}\left(dW_1^{Q1}(t), dW_2^{Q2}(t)\right) \label{corr}
\end{align}
Equation \eqref{corr} holds under both measure $Q1$ and $Q2$; thus, Girsanov transform under the change of measure preserves the instantaneous correlation between Brownian motions.

Let us now formally introduce G2++ and Heston models, that will be used to demonstrate our estimation methodology.

\textbf{G2++}
\begin{gather}
r_t = \phi(t) + x_t + y_t \\
dx_t = -a x_t dt + \sigma dW_t^x, \ x_0 = 0 \\
dy_t = -b y_t dt + \eta dW_t^y, \ y_0 = 0 \\
dW_t^xdW_t^y = \rho_{x,y} \ dt
\end{gather}

$r_t$ is the short rate, $x_t$, $y_t$ are some unobservable states, $\phi(t)$ is a deterministic function used to match the term structure at time 0, $a$, $b$, $\sigma$, $\eta$ are strictly positive constants, and $\rho_{x,y}$ is the instantaneous correlation between the Brownian motions $dW_t^x$ and $dW_t^y$

\textbf{Heston}
\begin{gather}
ds_t = (\tilde{r}_t-\tilde{q}_t-v_t/2)dt + \sqrt{v_t} dW_t^s \label{heston1}\\
dv_t = \kappa(\theta - v_t) dt + \xi \sqrt{v_t} dW_t^v \label{heston2} \\
dW_t^s dW_t^v = \rho_{s,v} \ dt \label{heston3}
\end{gather}

$s_t = \log(S_t)$ is the natural log of the stock price $S_t$, $v_t$ is the unobservable stochastic variance, $\tilde{r}_t, \tilde{q}_t$ are deterministic short rate and dividend yield respectively, $\kappa$, $\theta$, $\xi$, $v_0$ are strictly positive constants and $\rho_{s,v}$ is the instantaneous correlation between the Brownian motions $dW_t^s$ and $dW_t^v$.

\section{Correlation Estimation} \label{sec:est}

The component models in the hybrid system will be indexed by the subscripts $\{1,2,...\}$, and each component can be either a G2++ or Heston. G2++ is commonly used to describe the short rate process while equity or exchange rate is often modeled by Heston.

If the $i^{th}$ component is a G2++, $(x_i(t), y_i(t))$ will be the corresponding state vector, otherwise the Heston model will have states $(s_i(t), v_i(t))$. The instantaneous correlation between the Brownian motions $dW_i^p(t), dW_j^q(t)$ of states $p_i(t)$ and $q_j(t)$ will be denoted as $\rho_{p_i,q_j}$, where $p_i, q_j$ can be any state from G2++ or Heston.

The diagonal blocks of the instantaneous correlation matrix are assumed to be calibrated by the observed volatility surfaces/cubes.

\subsection{G2++/G2++}

In this section, both the $i^{th}$ and $j^{th}$ models are assumed to be G2++. Although the states $x_t$, $y_t$ of G2++ are not observable, we will show that correlations can be inferred from the spot rates. Let $R_i^{\tau_p}(t)$ and $R_j^{\tau_q}(t)$ be the continuously compounding spot rate of the $i^{th}$ and $j^{th}$ component with tenor $\tau_p$ and $\tau_q$ at time $t$ respectively. It is supposed that they are observed at times $\{t_0=0,t_1,t_2,...,t_n=T\}$.  $\Delta^{R}_{i,\tau}(t_k)$ is defined as

\begin{gather}
	\Delta^{R}_{i,\tau}(k) \equiv R_i^{\tau}(t_k) - R_i^{\tau}(t_{k-1})
\end{gather}

The empirical correlation $\hat{\rho}_{n,T}(\Delta^{R}_{i,\tau_p},\Delta^{R}_{j,\tau_q})$ is defined as the sample correlation between $\Delta^{R}_{i,\tau_p}$ and $\Delta^{R}_{j,\tau_q}$
\begin{gather}
\hat{\rho}_{n,T}(\Delta^{R}_{i,\tau_p},\Delta^{R}_{j,\tau_q}) \equiv \frac{ \sum_{k=1}^n (\Delta^{R}_{i,\tau_p}(k) - \overline{\Delta}^{R}_{i,\tau_p})(\Delta^{R}_{j,\tau_q}(k) - \overline{\Delta}^{R}_{j,\tau_q}) }
	{\sqrt{{\sum_{k=1}^n (\Delta^{R}_{i,\tau_p}(k) - \overline{\Delta}^{R}_{i,\tau_p})^2} {\sum_{k=1}^n (\Delta^{R}_{j,\tau_q}(k) - \overline{\Delta}^{R}_{j,\tau_q})^2}}} \\
	\overline{\Delta}^{R}_{i,\tau_p} = \sum_{k=1}^n \frac{\Delta^{R}_{i,\tau_p}(k)}{n}, \quad \overline{\Delta}^{R}_{j,\tau_q} = \sum_{k=1}^n \frac{\Delta^{R}_{j,\tau_q}(k)}{n}
\end{gather}

\begin{thm} \label{thm1}
\begin{gather}
\hat{\rho}_{n,T}(\Delta^{R}_{i,\tau_p},\Delta^{R}_{j,\tau_q}) \xrightarrow[p]{\quad \ n \rightarrow \infty \quad} 
c_{1,1}(i,j,\tau_p,\tau_q) \rho_{x_i,x_j} + c_{1,2}(i,j,\tau_p,\tau_q) \rho_{x_i,y_j} \notag \\
+ c_{2,1}(i,j,\tau_p,\tau_q) \rho_{y_i,x_j} + c_{2,2}(i,j,\tau_p,\tau_q) \rho_{y_i,y_j}
\end{gather}
where,
\begin{gather}
c_{1,1}(i,j,\tau_p,\tau_q) = \frac{ c(a_i,\sigma_i,\tau_p)c(a_j,\sigma_j,\tau_q) } 
{ d(a_i,b_i,\sigma_i,\eta_i,\tau_p,\rho_{x_i,y_i}) d(a_j,b_j,\sigma_j,\eta_j,\tau_q,\rho_{x_j,y_j})}\\
c_{1,2}(i,j,\tau_p,\tau_q) = \frac{ c(a_i,\sigma_i,\tau_p)c(b_j,\eta_j,\tau_q) } 
{ d(a_i,b_i,\sigma_i,\eta_i,\tau_p,\rho_{x_i,y_i}) d(a_j,b_j,\sigma_j,\eta_j,\tau_q,\rho_{x_j,y_j})}\\
c_{2,1}(i,j,\tau_p,\tau_q) = \frac{ c(b_i,\eta_i,\tau_p)c(a_j,\sigma_j,\tau_q)  }  
{ d(a_i,b_i,\sigma_i,\eta_i,\tau_p,\rho_{x_i,y_i}) d(a_j,b_j,\sigma_j,\eta_j,\tau_q,\rho_{x_j,y_j})}\\
c_{2,2}(i,j,\tau_p,\tau_q) = \frac{ c(b_i,\eta_i,\tau_p)c(b_j,\eta_j,\tau_q) } 
{ d(a_i,b_i,\sigma_i,\eta_i,\tau_p,\rho_{x_i,y_i}) d(a_j,b_j,\sigma_j,\eta_j,\tau_q,\rho_{x_j,y_j})} \\
d(\lambda_1,\lambda_2, \gamma_1, \gamma_2, \tau, \rho) = \sqrt{ c(\lambda_1,\gamma_1,\tau)^2  + c(\lambda_2,\gamma_2,\tau)^2 
+ 2c(\lambda_1,\gamma_1,\tau)c(\lambda_2,\gamma_2,\tau)\rho } \\
c(\lambda,\gamma,\tau) = \frac{ \gamma (1 - \exp(-\lambda \tau)) } { \lambda \tau}
\end{gather}
\end{thm}
Remark: The notation of $\xrightarrow[p]{\quad \ n \rightarrow \infty \quad}$ means convergence in probability $(p)$ as the number of time steps $n$ goes to $\infty$, while keeping the terminal date $T$ fixed.
\begin{proof}
The bond price under G2++ is given by \parencite{14}
\begin{equation}
	P(t,t+\tau) = \exp \left\{ -\int_t^{t+\tau} \phi(s)ds + \frac{V(t,t+\tau)}{2}  - \frac{(1-\e^{-a\tau})x_t}{a} - \frac{(1-\e^{-b\tau})y_t}{b}  \right\}
\end{equation}
where $V(t,t+\tau)$ is a finite variation function. Using the well-known result of the solution of Ornstein–Uhlenbeck (OU) process, the continuously compounding spot rate is
\begin{align}
	R^\tau(t) &= \frac{- \log(P_{t,t+\tau})}{\tau} = \frac{\int_t^{t+\tau} \phi(s)ds}{\tau} - \frac{V(t,t+\tau)}{2\tau}  + \frac{(1-\e^{-a\tau})x_t}{a\tau} + \frac{(1-\e^{-b\tau})y_t}{b\tau}  \notag \\
	&= A_t + \frac{1-\e^{-a\tau}}{a\tau} \int_0^t \sigma dW_u^x + \frac{1-\e^{-b\tau}}{b\tau} \int_0^t \eta dW_u^y \notag  \\
	&= A_t + c(a,\sigma,\tau) W_t^x + c(b,\eta,\tau) W_t^y
\end{align}
where $A_t$ is a finite variation term not used in later calculations. A simple transformation of the empirical correlation yields
\begin{equation}
	\hat{\rho}_{n,T}(\Delta^{R}_{i,\tau_p},\Delta^{R}_{j,\tau_q}) = \frac{\sum_{k=1}^n \Delta^{R}_{i,\tau_p}(k)\Delta^{R}_{j,\tau_q}(k) - n\overline{\Delta}^{R}_{i,\tau_p}\overline{\Delta}^{R}_{j,\tau_q}}
	{\sqrt{ \Bigl(\sum_{k=1}^n (\Delta^{R}_{i,\tau_p}(k))^2 - n (\overline{\Delta}^{R}_{i,\tau_p})^2 \Bigr) \Bigl( \sum_{k=1}^n (\Delta^{R}_{j,\tau_q}(k))^2 - n (\overline{\Delta}^{R}_{j,\tau_q})^2 \Bigr) } } 
\end{equation}
By the definition of quadratic covariation,
\begin{gather}
	\sum_{k=1}^n \Delta^{R}_{i,\tau_p}(k)\Delta^{R}_{j,\tau_q}(k) = \sum_{k=1}^n (R_i^{\tau_p}(t_k) - R_i^{\tau_p}(t_{k-1}))(R_j^{\tau_q}(t_k) - R_j^{\tau_q}(t_{k-1})) 
	\xrightarrow[p]{\quad n \rightarrow \infty \quad} [R_i^{\tau_p},R_j^{\tau_q}]_T
\end{gather}
Using the property of the quadratic covariation of Ito process,
\begin{flalign}
& [ R_i^{\tau_p},R_j^{\tau_q}]_T \notag \\
&= \Biggl[ c(a_i,\sigma_i,\tau_p) W^{x_i}_\bullet + c(b_i,\eta_i,\tau_p) W^{y_i}_\bullet, 
  c(a_j,\sigma_j,\tau_q) W^{x_j}_\bullet + c(b_j,\eta_j,\tau_q) W^{y_j}_\bullet \Biggr]_T & \notag \\
&= \Bigl( c(a_i,\sigma_i,\tau_p)c(a_j,\sigma_j,\tau_q) \rho_{x_i,x_j} + c(a_i,\sigma_i,\tau_p)c(b_j,\eta_j,\tau_q) \rho_{x_i,y_j}  \notag &\\
& \qquad \quad \quad + c(b_i,\eta_i,\tau_p)c(a_j,\sigma_j,\tau_q) \rho_{y_i,x_j} + c(b_i,\eta_i,\tau_p)c(b_j,\eta_j,\tau_q) \rho_{y_i,y_j} \Bigr)T & 
\end{flalign}

Similarly,
\begin{equation}
\sum_{k=1}^n \Delta^{R}_{i,\tau_p}(k)^2 	\xrightarrow[p]{\quad n \rightarrow \infty \quad} 
\Bigl( c(a_i,\sigma_i,\tau_p)^2 + c(b_i,\eta_i,\tau_p)^2 
+ 2\rho_{x_i,y_i}c(a_i,\sigma_i,\tau_p) c(b_i,\eta_i,\tau_p) \Bigr)  T \\
\end{equation}
\begin{equation}
\sum_{k=1}^n \Delta^{R}_{j,\tau_q}(k)^2 	\xrightarrow[p]{\quad n \rightarrow \infty \quad} 
\Bigl( c(a_j,\sigma_j,\tau_q)^2  + c(b_j,\eta_j,\tau_q)^2 
+ 2\rho_{x_j,y_j}c(a_j,\sigma_j,\tau_q) c(b_j,\eta_j,\tau_q)  \Bigr) T 
\end{equation}
Observing that $R_i^\tau(T) \in L^2$, for any $\varepsilon$ $>$ 0, by Markov inequality
\begin{equation}
	\mathbb{P} \left( \frac{|R_i^\tau(T)|}{\sqrt{n}} \ge \varepsilon \right) \le \frac{\E(R_i^\tau(T)^2)} { n \varepsilon^2} \xrightarrow{\quad n \rightarrow \infty \quad}  0
\end{equation}
Therefore,
\begin{flalign}
	n \overline{\Delta}^{R}_{i,\tau_p} \overline{\Delta}^{R}_{j,\tau_q} 
	&= \frac{ \Bigl( \sum_{k=1}^n \Delta^{R}_{i,\tau_p}(k) \Bigr) \Bigl( \sum_{k=1n}^n \Delta^{R}_{j,\tau_q}(k) \Bigr)}{n} & \notag \\ 
	&= \frac{ (R_i^{\tau_p}(T) - R_i^{\tau_p}(0)) } {\sqrt{n}} \frac{(R_j^{\tau_q}(T) - R_j^{\tau_q}(0)) } { \sqrt{n}}
	\xrightarrow[p]{\quad n \rightarrow \infty \quad} 0 &
\end{flalign}
\begin{flalign}
	& n (\overline{\Delta}^{R}_{i,\tau_p})^2  \xrightarrow[p]{\quad n \rightarrow \infty \quad} 0, \quad
	n (\overline{\Delta}^{R}_{j,\tau_q})^2  \xrightarrow[p]{\quad n \rightarrow \infty \quad} 0 &
\end{flalign}
Combining all the components, result follows.
\end{proof}

In view of theorem \ref{thm1}, if ones can observe two distinct spot rates for each component model\footnote{Similarly, we need $n_1$ and $n_2$ distinct spot rates for the $Gn_1++$/$Gn_2++$ combination respectively, and the linear system will be $n_1n_2 \times n_1n_2$.}, they can back out the instantaneous correlations from the empirical ones by solving the following 4x4 linear system.

\begin{equation}
	A\begin{bmatrix}
		\rho_{x_i,x_j} \\
		\rho_{x_i,y_j}\\
		\rho_{y_i,x_j}\\
		\rho_{y_i,y_j} \\
	\end{bmatrix}
	= \begin{bmatrix}
		\hat{\rho}_{n,T}(\Delta^R_{i,\tau_{p1}}, \Delta^R_{j,\tau_{q1}}) \\
		\hat{\rho}_{n,T}(\Delta^R_{i,\tau_{p1}}, \Delta^R_{j,\tau_{q2}}) \\
		\hat{\rho}_{n,T}(\Delta^R_{i,\tau_{p2}}, \Delta^R_{j,\tau_{q1}}) \\	
		\hat{\rho}_{n,T}(\Delta^R_{i,\tau_{p2}}, \Delta^R_{j,\tau_{q2}}) \\	
	\end{bmatrix}
\end{equation}
where
\begin{equation}
	A=\begin{bmatrix}
		c_{1,1}(i,j,\tau_{p1},\tau_{q1}) & c_{1,2}(i,j,\tau_{p1},\tau_{q1}) & c_{2,1}(i,j,\tau_{p1},\tau_{q1}) & c_{2,2}(i,j,\tau_{p1},\tau_{q1})\\
		c_{1,1}(i,j,\tau_{p1},\tau_{q2}) & c_{1,2}(i,j,\tau_{p1},\tau_{q2}) & c_{2,1}(i,j,\tau_{p1},\tau_{q2}) & c_{2,2}(i,j,\tau_{p1},\tau_{q2})\\
		c_{1,1}(i,j,\tau_{p2},\tau_{q1}) & c_{1,2}(i,j,\tau_{p2},\tau_{q1}) & c_{2,1}(i,j,\tau_{p2},\tau_{q1}) & c_{2,2}(i,j,\tau_{p2},\tau_{q1})\\
		c_{1,1}(i,j,\tau_{p2},\tau_{q2}) & c_{1,2}(i,j,\tau_{p2},\tau_{q2}) & c_{2,1}(i,j,\tau_{p2},\tau_{q2}) & c_{2,2}(i,j,\tau_{p2},\tau_{q2})\\
	\end{bmatrix}
\end{equation}

%\begin{small}
%\begin{equation}
%\begin{bmatrix}
%	c_{1,1}(i,j,\tau_{p1},\tau_{q1}) & c_{1,2}(i,j,\tau_{p1},\tau_{q1}) & c_{2,1}(i,j,\tau_{p1},\tau_{q1}) & c_{2,2}(i,j,\tau_{p1},\tau_{q1})\\
%	c_{1,1}(i,j,\tau_{p1},\tau_{q2}) & c_{1,2}(i,j,\tau_{p1},\tau_{q2}) & c_{2,1}(i,j,\tau_{p1},\tau_{q2}) & c_{2,2}(i,j,\tau_{p1},\tau_{q2})\\
%	c_{1,1}(i,j,\tau_{p2},\tau_{q1}) & c_{1,2}(i,j,\tau_{p2},\tau_{q1}) & c_{2,1}(i,j,\tau_{p2},\tau_{q1}) & c_{2,2}(i,j,\tau_{p2},\tau_{q1})\\
%	c_{1,1}(i,j,\tau_{p2},\tau_{q2}) & c_{1,2}(i,j,\tau_{p2},\tau_{q2}) & c_{2,1}(i,j,\tau_{p2},\tau_{q2}) & c_{2,2}(i,j,\tau_{p2},\tau_{q2})\\
%\end{bmatrix}
%\begin{bmatrix}
%	\rho_{x_i,x_j} \\
%	\rho_{x_i,y_j}\\
%	\rho_{y_i,x_j}\\
%	\rho_{y_i,y_j} \\
%\end{bmatrix}
%=
%\begin{bmatrix}
%	\hat{\rho}_{n,T}(\Delta^R_{i,\tau_{p1}}, \Delta^R_{j,\tau_{q1}}) \\
%	\hat{\rho}_{n,T}(\Delta^R_{i,\tau_{p1}}, \Delta^R_{j,\tau_{q2}}) \\
%	\hat{\rho}_{n,T}(\Delta^R_{i,\tau_{p2}}, \Delta^R_{j,\tau_{q1}}) \\	
%	\hat{\rho}_{n,T}(\Delta^R_{i,\tau_{p2}}, \Delta^R_{j,\tau_{q2}}) \\	
%\end{bmatrix}
%\end{equation}
%\end{small}

\subsection{G2++/Heston}

In the Heston model, the stochastic variance $v_t$ is not observable. Hence, as in \textcite{11}, the \textit{square} of the short-term at-the-money implied volatility is used as a proxy for $v_t$, justified by the small-time asymptotics of Heston \parencite{10}.\\

\begin{thm}[\textcite{10}] Let $V(K,T)$ be the implied volatility of a European call option with strike K and maturity T at time 0. If $S_t$ follows the Heston process \eqref{heston1} - \eqref{heston3}, then
	\begin{equation}
		\lim_{T \rightarrow 0} V(K,T) = \sqrt{v_0} \Bigl[ 1 + \frac{\rho_{s,v} z}{4} + \Bigl(\frac{1}{24} - \frac{5 \rho^2}{48} \Bigr)z^2 + O(z^3) \Bigr]
	\end{equation}
where $z = \frac{\xi \log(K/S_0)}{v_0}$
\end{thm}

In this section, it is assumed that the log stock price $s_j(t)$ and stochastic variance $v_j(t)$ in the $j^{th}$ component are observed at times $\{t_0=0,t_1,t_2,...,t_n=T\}$, and $\Delta^s_j(t_k)$ and $\Delta^v_j(t_k)$ are defined as

\begin{gather}
	\Delta^s_j(k) \equiv s_j(t_k) - s_j(t_{k-1}) \\
	\Delta^v_j(k) \equiv v_j(t_k) - v_j(t_{k-1})
\end{gather}

The empirical correlation $\hat{\rho}_{n,T}(\Delta^{R}_{i,\tau},\Delta^{s}_{j})$ is defined as the sample correlation between $\Delta^{R}_{i,\tau}$ and $\Delta^{s}_{j}$
\begin{gather}
	\hat{\rho}_{n,T}(\Delta^{R}_{i,\tau},\Delta^{s}_{j}) \equiv \frac{ \sum_{k=1}^n (\Delta^{R}_{i,\tau}(k) - \overline{\Delta}^{R}_{i,\tau})(\Delta^{s}_{j}(k) - \overline{\Delta}^{s}_{j}) }
	{\sqrt{{\sum_{k=1}^n (\Delta^{R}_{i,\tau}(k) - \overline{\Delta}^{R}_{i,\tau})^2} {\sum_{k=1}^n (\Delta^{s}_{j}(k) - \overline{\Delta}^{s}_{j})^2}}} \\
	\overline{\Delta}^{R}_{i,\tau} = \sum_{k=1}^n \frac{\Delta^{R}_{i,\tau}(k)}{n}, \quad \overline{\Delta}^{s}_{i} = \sum_{k=1}^n \frac{\Delta^{s}_{i}(k)}{n}
\end{gather}
Similarly, for $\hat{\rho}_{n,T}(\Delta^{R}_{i,\tau},\Delta^{v}_{j})$
\begin{gather}
	\hat{\rho}_{n,T}(\Delta^{R}_{i,\tau},\Delta^{v}_{j}) \equiv \frac{ \sum_{k=1}^n (\Delta^{R}_{i,\tau}(k) - \overline{\Delta}^{R}_{i,\tau})(\Delta^{v}_{j}(k) - \overline{\Delta}^{v}_{j}) }
	{\sqrt{{\sum_{k=1}^n (\Delta^{R}_{i,\tau}(k) - \overline{\Delta}^{R}_{i,\tau})^2} {\sum_{k=1}^n (\Delta^{v}_{j}(k) - \overline{\Delta}^{v}_{j})^2}}} \\
	\overline{\Delta}^{R}_{i,\tau} = \sum_{k=1}^n \frac{\Delta^{R}_{i,\tau}(k)}{n}, \quad \overline{\Delta}^{v}_{i} = \sum_{k=1}^n \frac{\Delta^{v}_{i}(k)}{n}
\end{gather}

\begin{thm} \label{thm2}
	\begin{gather}
		\hat{\rho}_{n,T}(\Delta^{R}_{i,\tau},\Delta^{s}_{j}) \xrightarrow[p]{\quad \ T \rightarrow 0, \ n \rightarrow \infty \quad} 
		c_{1}(i,\tau) \rho_{x_i,s_j} + c_{2}(i,\tau) \rho_{y_i,s_j} \label{bs_only} \\
		\hat{\rho}_{n,T}(\Delta^{R}_{i,\tau},\Delta^{v}_{j}) \xrightarrow[p]{\quad \ T \rightarrow 0, \ n \rightarrow \infty \quad} 
		c_{1}(i,\tau) \rho_{x_i,v_j} + c_{2}(i,\tau) \rho_{y_i,v_j}
	\end{gather}
	where,
	\begin{gather}
		c_{1}(i,\tau) = \frac{ c(a_i,\sigma_i,\tau) } 
		{ d(a_i,b_i,\sigma_i,\eta_i,\tau,\rho_{x_i,y_i}) }\\
		c_{2}(i,\tau) = \frac{ c(b_i,\eta_i,\tau)  }  
		{ d(a_i,b_i,\sigma_i,\eta_i,\tau,\rho_{x_i,y_i}) }\\
		d(\lambda_1,\lambda_2, \gamma_1, \gamma_2, \tau, \rho) = \sqrt{ c(\lambda_1,\gamma_1,\tau)^2  + c(\lambda_2,\gamma_2,\tau)^2 
		+ 2c(\lambda_1,\gamma_1,\tau)c(\lambda_2,\gamma_2,\tau)\rho } \\
		c(\lambda,\gamma,\tau) = \frac{ \gamma (1 - \exp(-\lambda \tau)) } { \lambda \tau}
	\end{gather}
\end{thm}
Remark: The notation of $\xrightarrow[p]{\quad \ T \rightarrow 0, \ n \rightarrow \infty \quad}$ means convergence in probability $(p)$ as the number of time steps $n$ goes to $\infty$, and the terminal date $T$ go to 0. No joint convergence is needed, we only require iterative convergence of first $n$ goes to $\infty$ with $T$ fixed, and then $T$ goes to 0.
\begin{proof}
	\begin{equation}
		\hat{\rho}_{n,T}(\Delta^{R}_{i,\tau},\Delta^{s}_{j}) = \frac{\sum_{k=1}^n \Delta^{R}_{i,\tau}(k)\Delta^{s}_{j}(k) - n\overline{\Delta}^{R}_{i,\tau}\overline{\Delta}^{s}_{j}}
		{\sqrt{ \Bigl(\sum_{k=1}^n (\Delta^{R}_{i,\tau}(k))^2 - n (\overline{\Delta}^{R}_{i,\tau})^2 \Bigr) \Bigl( \sum_{k=1}^n (\Delta^{s}_{j}(k))^2 - n (\overline{\Delta}^{s}_{j})^2 \Bigr) } } 
	\end{equation}
	For a given $T>0$, by the definition of quadratic covariation
	\begin{gather}
		\sum_{k=1}^n \Delta^{R}_{i,\tau}(k)\Delta^{s}_{j}(k) = \sum_{k=1}^n (R_i^{\tau}(t_k) - R_i^{\tau}(t_{k-1}))(s_j(t_k) - s_j(t_{k-1})) 
		\xrightarrow[p]{\quad n \rightarrow \infty \quad} [R_i^{\tau},s_j]_T
	\end{gather}
	Using the well-known result for the quadratic covariation of Ito process,
	\begin{align}
	[ R_i^{\tau},s_j]_T 
	&= \Biggl[ c(a_i,\sigma_i,\tau) W^{x_i}_\bullet + c(b_i,\eta_i,\tau) W^{y_i}_\bullet,
		  \int_0^\bullet \sqrt{v_j(t)} dW_t^{s_j} \Biggr]_T \notag \\
	&=   c(a_i,\sigma_i,\tau) \rho_{x_i,s_j} \int_0^T \sqrt{v_j(t)} dt 
		 + c(b_i,\eta_i,\tau) \rho_{y_i,s_j}  \int_0^T \sqrt{v_j(t)}  dt 
	\end{align}
	Similarly,
	\begin{gather}
		\sum_{k=1}^n \Delta^{R}_{i,\tau}(k)^2 	\xrightarrow[p]{\quad n \rightarrow \infty \quad} 
		\Bigl( c(a_i,\sigma_i,\tau)^2 + c(b_i,\eta_i,\tau)^2 
		+ 2\rho_{x_i,y_i}c(a_i,\sigma_i,\tau)c(b_i,\eta_i,\tau) \Bigr)  T  \\
		\sum_{k=1}^n \Delta^{s}_{j}(k)^2 \xrightarrow[p]{\quad n \rightarrow \infty \quad} 
		\int_0^T v_j(t) dt 
	\end{gather}
	Following the same argument as theorem \ref{thm1}
	\begin{gather}
		n (\overline{\Delta}^{R}_{i,\tau})^2  \xrightarrow[p]{\quad n \rightarrow \infty \quad} 0, \quad
		n \overline{\Delta}^{R}_{j,\tau_q}\overline{\Delta}^{s}_{j}  \xrightarrow[p]{\quad n \rightarrow \infty \quad} 0, \quad
		n (\overline{\Delta}^{s}_{j})^2	\xrightarrow[p]{\quad n \rightarrow \infty \quad} 0
	\end{gather}
	Finally, since $v_j(t)$ is continuous,
	\begin{gather}
		\lim_{T \rightarrow 0} \frac{\int_0^T \sqrt{v_j(t)} dt}{T} = \sqrt{v_j(0)}, 
		\quad \lim_{T \rightarrow 0} \frac{\int_0^T v_j(t) dt}{T} = v_j(0)
	\end{gather}
	Combining all the components, result for $\hat{\rho}_{n,T}(\Delta^{R}_{i,\tau},\Delta^{s}_{j})$ follows, and the arguments for $\hat{\rho}_{n,T}(\Delta^{R}_{i,\tau},\Delta^{v}_{j})$ are similar.
\end{proof}

Unlike the case of G2++/G2++, the empirical correlation only converges to a constant when $T$ tends to 0, but we will show numerically in Section \ref{sec:num} that the error due to finite $T$ is insignificant.

By computing the empirical correlation of two distinct spot rates with the log stock price and the variance proxy, one can obtain the instantaneous correlation via the following block diagonal linear system:

\begin{equation}
	\begin{bmatrix}
		c_{1}(i,\tau_{1}) & c_{2}(i,\tau_{1}) & 0 & 0\\
		c_{1}(i,\tau_{2}) & c_{2}(i,\tau_{2}) & 0 & 0\\
		0 & 0 & c_{1}(i,\tau_{1}) & c_{2}(i,\tau_{1})\\
		0 & 0 & c_{1}(i,\tau_{2}) & c_{2}(i,\tau_{2})\\
	\end{bmatrix}
	\begin{bmatrix}
		\rho_{x_i,s_j} \\
		\rho_{y_i,s_j}\\
		\rho_{x_i,v_j}\\
		\rho_{y_i,v_j} \\
	\end{bmatrix}
	=
	\begin{bmatrix}
		\hat{\rho}_{n,T}(\Delta^R_{i,\tau_{1}}, \Delta^s_{j}) \\
		\hat{\rho}_{n,T}(\Delta^R_{i,\tau_{2}}, \Delta^s_{j}) \\
		\hat{\rho}_{n,T}(\Delta^R_{i,\tau_{1}}, \Delta^v_{j}) \\	
		\hat{\rho}_{n,T}(\Delta^R_{i,\tau_{2}}, \Delta^v_{j}) \\	
	\end{bmatrix}
\end{equation}
%This block diagonal system can be easily solved in closed form.
%\begin{gather}
%	\rho_{x_i,s_j} =\frac{ c_2(i,\tau_2) \hat{\rho}_{n,T}(\Delta^R_{i,\tau_{1}}, \Delta^s_{j}) 
%		- c_2(i,\tau_1) \hat{\rho}_{n,T}(\Delta^R_{i,\tau_{2}}, \Delta^s_{j}) }
%	{c_1(i,\tau_1)c_2(i,\tau_2) - c_1(i,\tau_2)c(i,\tau_1)} \\
%	\rho_{y_i,s_j} =\frac{ c_1(i,\tau_1) \hat{\rho}_{n,T}(\Delta^R_{i,\tau_{2}}, \Delta^s_{j}) 
%		- c_1(i,\tau_2) \hat{\rho}_{n,T}(\Delta^R_{i,\tau_{1}}, \Delta^s_{j})}
%	{c_1(i,\tau_1)c_2(i,\tau_2) - c_1(i,\tau_2)c(i,\tau_1)} \\
%	\rho_{x_i,v_j} =\frac{ c_2(i,\tau_2) \hat{\rho}_{n,T}(\Delta^R_{i,\tau_{1}}, \Delta^v_{j}) 
%	- c_2(i,\tau_1) \hat{\rho}_{n,T}(\Delta^R_{i,\tau_{2}}, \Delta^v_{j}) }
%	{c_1(i,\tau_1)c_2(i,\tau_2) - c_1(i,\tau_2)c(i,\tau_1)} \\
%	\rho_{y_i,v_j} =\frac{ c_1(i,\tau_1) \hat{\rho}_{n,T}(\Delta^R_{i,\tau_{2}}, \Delta^v{j}) 
%	- c_1(i,\tau_2) \hat{\rho}_{n,T}(\Delta^R_{i,\tau_{1}}, \Delta^v_{j})}
%	{c_1(i,\tau_1)c_2(i,\tau_2) - c_1(i,\tau_2)c(i,\tau_1)} 
%\end{gather}

\subsection{Heston/Heston}

The Heston/Heston case is even simpler, when $n$ goes to infinity and $T$ goes to 0, the empirical correlations converge directly to the instantaneous correlation.

\begin{thm} \label{thm3}
	\begin{gather}
		\hat{\rho}_{n,T}(\Delta^{s}_{i},\Delta^{s}_{j}) \xrightarrow[p]{\quad \ T \rightarrow 0, \ n \rightarrow \infty \quad} \rho_{s_i,s_j}, \quad 		\hat{\rho}_{n,T}(\Delta^{s}_{i},\Delta^{v}_{j}) \xrightarrow[p]{\quad \ T \rightarrow 0, \ n \rightarrow \infty \quad} \rho_{s_i,v_j} \\
		\hat{\rho}_{n,T}(\Delta^{v}_{i},\Delta^{s}_{j}) \xrightarrow[p]{\quad \ T \rightarrow 0, \ n \rightarrow \infty \quad} \rho_{v_i,s_j}, \quad 
		\hat{\rho}_{n,T}(\Delta^{v}_{i},\Delta^{v}_{j}) \xrightarrow[p]{\quad \ T \rightarrow 0, \ n \rightarrow \infty \quad} \rho_{v_i,v_j}
	\end{gather}
\end{thm}
\begin{proof}
	\begin{equation}
		\hat{\rho}_{n,T}(\Delta^{s}_{i},\Delta^{s}_{j}) = \frac{\sum_{k=1}^n \Delta^{s}_{i}(k)\Delta^{s}_{j}(k) - n\overline{\Delta}^{s}_{i}\overline{\Delta}^{s}_{j}}
		{\sqrt{ \Bigl(\sum_{k=1}^n (\Delta^{s}_{i}(k))^2 - n (\overline{\Delta}^{s}_{i})^2 \Bigr) \Bigl( \sum_{k=1}^n (\Delta^{s}_{j}(k))^2 - n (\overline{\Delta}^{s}_{j})^2 \Bigr) } } 
	\end{equation}
	For a given $T>0$, by the definition of quadratic covariation
	\begin{gather}
		\sum_{k=1}^n \Delta^{s}_{i}(k)\Delta^{s}_{j}(k) = \sum_{k=1}^n (s_i(t_k) - s_i(t_{k-1}))(s_j(t_k) - s_j(t_{k-1})) 
		\xrightarrow[p]{\quad n \rightarrow \infty \quad} [s_i,s_j]_T
	\end{gather}
	\begin{equation}
		[ s_i,s_j]_T 
		= \Biggl[\int_0^\bullet \sqrt{v_i(t)} dW_t^{s_i}, \int_0^\bullet \sqrt{v_j(t)} dW_t^{s_j} \Biggr]_T 
		=  \rho_{s_i,s_j}\int_0^T  \sqrt{v_i(t)v_j(t)}  dt  
	\end{equation}
	Similarly,
	\begin{gather}
		\sum_{k=1}^n \Delta^{s}_{i}(k)^2 	\xrightarrow[p]{\quad n \rightarrow \infty \quad} 
		\int_0^T  v_i(t) dt, \
		\sum_{k=1}^n \Delta^{s}_{j}(k)^2 	\xrightarrow[p]{\quad n \rightarrow \infty \quad} 
		\int_0^T  v_j(t)dt 
	\end{gather}
	Following the same argument as theorem \ref{thm1}
	\begin{gather}
		n (\overline{\Delta}^{s}_{i})^2  \xrightarrow[p]{\quad n \rightarrow \infty \quad} 0, \quad
		n \overline{\Delta}^{s}_{i}\overline{\Delta}^{s}_{j}  \xrightarrow[p]{\quad n \rightarrow \infty \quad} 0, \quad
		n (\overline{\Delta}^{s}_{j})^2	\xrightarrow[p]{\quad n \rightarrow \infty \quad} 0 
	\end{gather}
	Finally, since $v_i(t), v_j(t)$ are continuous, 
	\begin{gather}
		\lim_{T \rightarrow 0} \frac{\int_0^T \sqrt{v_i(t)} dt}{T} = \sqrt{v_i(0)}, 
		\quad \lim_{T \rightarrow 0} \frac{\int_0^T v_i(t) dt}{T} = v_i(0)  \\
		\lim_{T \rightarrow 0} \frac{\int_0^T \sqrt{v_j(t)} dt}{T} = \sqrt{v_j(0)}, 
		\quad \lim_{T \rightarrow 0} \frac{\int_0^T v_j(t) dt}{T} = v_j(0)
	\end{gather}
	Combining all the components, result for $\hat{\rho}_{n,T}(\Delta^{s}_{i},\Delta^{s}_{j})$ follows. The arguments for other combinations are similar.
\end{proof}

\section{Missing Implied Volatility} \label{sec:miss}
Sometimes, the time series of implied volatility may not be available to compute the empirical correlations. There is a class of techniques called {\it positive semidefinite matrix completion} \parencite{25}, which can be used to fill in the missing values, but they are usually quite involved. Instead, a simple method that is used in \parencite{7} will be described here.

Let $W_i^1(t), W_i^2(t)$ be the correlated Brownian motions in the $i^{th}$ component model with correlation $\rho_i$, the de-correlated Brownian motion $Z_i^{1}(t), Z_i^{2}(t)$ is defined as
\begin{equation}
	\begin{bmatrix}
		Z_i^{1}(t) \\
		Z_i^{2}(t) \\
	\end{bmatrix}
	= L_i^{-1}
	\begin{bmatrix}
		W_i^{1}(t) \\
		W_i^{2}(t) \\
	\end{bmatrix}, \quad
	L_i = \text{chol}(\Sigma_i) =
	\begin{bmatrix}
		1 & 0 \\
		\rho_i & \sqrt{1 - \rho_i^2} \\
	\end{bmatrix},  \quad	
	\Sigma_i = 
	\begin{bmatrix}
		1 & \rho_i \\
		\rho_i & 1 \\
	\end{bmatrix}  \quad
\end{equation}
where chol($\bullet$) stands for the lower Choleskey decomposition. By construction, the correlation between $Z_i^{1}(t)$ and $Z_i^{2}(t)$ is zero.

Let $\Omega_{i,j}$ and $\Sigma_{i,j}$ be the cross-correlation matrix between $Z_i(t)'s, \ Z_j(t)'s$ and $W_i(t)'s, \ W_j(t)'s$ respectively, and it is easy to show that
\begin{gather} 
	\Omega_{i,j} = L_i^{-1} \Sigma_{i,j} (L_j^{-1})^T  \\
	\Sigma_{i,j} = L_i \Omega_{i,j} L_j^T \label{eq_sigma}
\end{gather}

\subsection{G2++/Heston}
For the case of G2++/Heston, it is assumed that the cross-correlation $\Omega_{i,j}$ has the form\\
\begin{equation}
	\Omega_{i,j}=
	\begin{bmatrix}
		c_{11} & 0\\
		c_{21} & 0 \\
	\end{bmatrix}
\end{equation}
Essentially, it means the correlations between the de-correlated variance and spot rates are zero. Using \eqref{eq_sigma}, the correlation matrix of $W_i(t)'s$ is given by

\begin{gather}
	\Sigma_{i,j} = L_i \Omega_{i,j} L_j^T =
	\begin{bmatrix}
		1 & 0 \\
		\rho_i & \sqrt{1 - \rho_i^2} \\
	\end{bmatrix}
	\begin{bmatrix}
		c_{11} & 0\\
		c_{21} & 0 \\
	\end{bmatrix}	
	\begin{bmatrix}
		1 & 0 \\
		\rho_j & \sqrt{1 - \rho_j^2} \\
	\end{bmatrix}^T	\notag \\
	= 	
	\begin{bmatrix}
		c_{11} & c_{11}\rho_j\\
		c_{11}\rho_i + c_{21}\sqrt{1-\rho_i^2} & \left( c_{11}\rho_i + c_{21}\sqrt{1-\rho_i^2} \right)\rho_j \\
	\end{bmatrix}	
\end{gather}
Since $\rho_{x_i,s_j} = c_{11}$ and $\rho_{y_i,s_j}=c_{11}\rho_i + c_{21}\sqrt{1-\rho_i^2}$, the instantaneous correlation for $v_j(t)$ can computed by
\begin{gather}
	\rho_{x_i,v_j} = \rho_{x_i,s_j} \rho_j\\
	\rho_{y_i,v_j} = \rho_{y_i,s_j} \rho_j
\end{gather}

The interpretation is that stochastic variance is coupled with spot rates via the stock. If the stock has strong linkage with the spot rates, so should be the variance. The strength and direction are then further modulated by the correlation between stock and variance.

\subsection{Heston/Heston}
If both components are Heston, there are two scenarios. If only one of the variances is missing, it will be similar to the case of G2++/Heston. Supposed both variances are missing, it is assumed that
\begin{equation}
	\Omega_{i,j}=
	\begin{bmatrix}
		c_{11} & 0\\
		0 & 0 \\
	\end{bmatrix}
\end{equation}
As a result,
\begin{gather}
	\Sigma_{i,j} = L_i \Omega_{i,j} L_j^T =
	\begin{bmatrix}
		1 & 0 \\
		\rho_i & \sqrt{1 - \rho_i^2} \\
	\end{bmatrix}
	\begin{bmatrix}
		c_{11} & 0\\
		0 & 0 \\
	\end{bmatrix}	
	\begin{bmatrix}
		1 & 0 \\
		\rho_j & \sqrt{1 - \rho_j^2} \\
	\end{bmatrix}^T	
	= 	
	\begin{bmatrix}
		c_{11} & c_{11}\rho_j\\
		c_{11}\rho_i & c_{11}\rho_i\rho_j \\
	\end{bmatrix}	
\end{gather}
Hence,
\begin{gather}
	\rho_{s_i,v_j} = \rho_{s_i,s_j} \rho_j\\
	\rho_{v_i,s_j} = \rho_{s_i,s_j} \rho_i\\
	\rho_{v_i,v_j} = \rho_{s_i,s_j} \rho_i \rho_j
\end{gather}

The interpretation is similar to that of G2++/Heston. All correlations are coupled indirectly via the stock $i$ and stock $j$ linkage, with strength and direction adjusted by the corresponding correlation between stock and variance.

\section{Transformation to Positive Semidefinite Matrix} \label{sec:trans}

Let $\hat{\Sigma}$ be a matrix with diagonal block correlations calibrated from the volatility surface and off-diagonal block entries estimated by empirical correlations using simultaneous equations. $\hat{\Sigma}$ cannot be a correlation matrix if it is not positive semidefinite. Furthermore, its entries coming from the system of linear equations may not even lie on [-1,1], when the model parameters are poorly fitted or the distribution implied by the volatility surface differs significantly from the historical distribution. 

To convert $\hat{\Sigma}$ to a correlation matrix, one way is to run an optimization to find the nearest positive semidefinite matrix according to the Frobenious norm \parencite{23,24}. 

To preserve the diagonal blocks implied by the volatility surface, one must introduce equality constraints, but optimizations with equality constraints are often hit-and-miss, as it is hard to find feasible candidates with equality constraints, especially in global optimizations, where the optimizers jump around to search for global optimum. If the initial guess is not feasible, the optimizer may be still moving around the infeasible region even after many iterations.

Other alternative approaches include alternating projection \parencite{22} and truncation of eigenvalues \parencite{21}, but they will change the entries in the diagonal blocks. 

In this paper, a recent approach called shrinking \parencite{4} is recommended. Given a symmetric indefinite matrix $M_0$ and a symmetric positive semidefinite matrix $M_1$, one wants to find a convex combination $S$ of $M_0$ and $M_1$ such that $S$ is positive semidefinite.
\begin{equation}
	S(\alpha) = (1-\alpha) M_0 + \alpha M_1
\end{equation}
$\alpha$ is called the shrinking parameter, and the minimum of $\alpha$ such that $S$ is positive semidefinite is called the optimal shrinking parameter $\alpha^*$. 
\begin{equation}
\alpha^* = \min \{ \alpha \in [0,1] \ | \ S(\alpha) \ \text{is positive semidefinite} \}
\end{equation}
$\alpha^*$ exist as $\alpha$=1 is always a solution. Furthermore, it can be shown that if $M_1$ is positive definite, the problem has an interior solution $\alpha^* \in (0,1)$. The authors have introduced three methods to compute $\alpha^*$ and the simplest bisection method is shown here.

\textbf{\textcite{4} Algorithm 3.1}
\begin{verbatim}
xl = 0, xr = 1
while (xr - xl > tol) {
    xm = (xl+xr)*0.5
    if ( S(xm) = (1-xm)*M0 + xm*M1 is not positive semidefinite ) {
      xl = xm
    } else {
      xr = xm
    }
}
return (1-xr)*M0 + xr*M1 
// need to use xr rather than xm to guarantee positive semidefiniteness
\end{verbatim}

$M_0$ corresponds to $\hat{\Sigma}$, but it is better to truncate the entries before shrinking, otherwise those large entries will force $\alpha^*$ to be very close to 1, resulting in no correlation for all off-diagonal blocks.

$M_1$ is the block diagonal matrix from the component model calibrations, with all off-diagonal blocks being 0. Provided that all diagonal blocks are positive definite, $M_1$ is also positive definite, thus $\alpha^* \in$ (0,1). Regardless of the value of $\alpha$, the diagonal blocks do not change, as they are the same in both $M_0$ and $M_1$. Compared with the constrained nearest positive semidefinite matrix optimization, the shrinking method is robust and extremely fast.

\section{Extensions} \label{sec:ext}

\subsection{Affine Term-Structure Models}

If the short rate model is affine \parencite{16}, then the spot rate can be expressed in the form
\begin{equation}
R_i^\tau(t) = \frac{A_i(t,t+\tau)}{\tau} + \sum_k \frac{B_{ik}(t,t+\tau) x_{ik}(t)}{\tau}
\end{equation}
where $A(t,T)$ and $B_{ik}(t,T)'s$ are deterministic differentiable functions. In addition, if $B_{ik}(t,T) = B_{ik}(T-t)$ and the states $\{x_{ik}(t)\}$ have drift coefficient $\{\mu_{ik}(t)\}$ and continuous\footnote{Drift and diffusion coefficients of SDE are very often continuous, otherwise strong solution may not exists. Thus, this requirement is not really restrictive.} diffusion coefficients $\{b_{ik}(t)\}$, for example, \textcite{30,31,32,33}, then
\begin{gather}
R_i^\tau(t) = \frac{A_i(t,t+\tau)}{\tau} + \sum_k \left[ \frac{B_{ik}(\tau)}{\tau}  \int_0^t \mu_{ik}(s) ds  \right] +
\sum_k \left[ \frac{B_{ik}(\tau)}{\tau}  \int_0^t b_{ik}(s) dW_s^{x_k}  \right] \\
[ R_i^{\tau_p},R_j^{\tau_q}]_T = \sum_k \sum_l \left[ \frac{B_{ik}(\tau)B_{jl}(\tau)\rho_{x_{ik},x_{jl}}}{\tau^2} \int_0^T b_{ik}(s)b_{jl}(s) ds \right] \\
[ R_i^{\tau},s_j]_T = \sum_k \left[ \frac{B_{ik}(\tau)\rho_{x_{ik},s_j}}{\tau} \int_0^T b_{ik}(s)\sqrt{v_j(s)} ds \right] \\
[ R_i^{\tau},v_j]_T = \sum_k \left[ \frac{B_{ik}(\tau)\xi \rho_{x_{ik},v_j}}{\tau} \int_0^T b_{ik}(s)\sqrt{v_j(s)} ds \right]
\end{gather}
It can be shown that theorem \ref{thm1} and \ref{thm2} can still be applied with the appropriate modifications.

\subsection{G1++}
Under G1++/G1++, we just need to pick one arbitrary spot rate for each component, and theorem \ref{thm1} degenerates such that the empirical correlation between spot rates converges to the instantaneous correlation.

\begin{equation}
\hat{\rho}_{n,T}(\Delta^{R}_{i,\tau_p},\Delta^{R}_{j,\tau_q}) \xrightarrow[p]{\quad \ n \rightarrow \infty \quad} 
\rho_{x_i,x_j}
\end{equation}

For the case of G1++/G2++, the 4x4 linear system becomes 2x2, with the coefficients modified as follows:
\begin{gather}
	\hat{\rho}_{n,T}(\Delta^{R}_{i,\tau_p},\Delta^{R}_{j,\tau_q}) \xrightarrow[p]{\quad \ n \rightarrow \infty \quad} 
	c_{1,1}(i,j,\tau_p,\tau_q) \rho_{x_i,x_j} + c_{1,2}(i,j,\tau_p,\tau_q) \rho_{x_i,y_j}
\end{gather}
where,
\begin{gather}
	c_{1,1}(i,j,\tau_p,\tau_q) = \frac{ c(a_j,\sigma_j,\tau_q) } 
	{ d(a_j,b_j,\sigma_j,\eta_j,\tau_q,\rho_{x_j,y_j})}\\
	c_{1,2}(i,j,\tau_p,\tau_q) = \frac{ c(b_j,\eta_j,\tau_q) } 
	{  d(a_j,b_j,\sigma_j,\eta_j,\tau_q,\rho_{x_j,y_j})}\\
	d(\lambda_1,\lambda_2, \gamma_1, \gamma_2, \tau, \rho) = \sqrt{ c(\lambda_1,\gamma_1,\tau)^2  + c(\lambda_2,\gamma_2,\tau)^2 
		+ 2c(\lambda_1,\gamma_1,\tau)c(\lambda_2,\gamma_2,\tau)\rho } \\
	c(\lambda,\gamma,\tau) = \frac{ \gamma (1 - \exp(-\lambda \tau)) } { \lambda \tau}
\end{gather}

Similarly, G1++/Heston will degenerate to give the instantaneous correlation directly.
\begin{gather}
	\hat{\rho}_{n,T}(\Delta^{R}_{i,\tau},\Delta^{s}_{j}) \xrightarrow[p]{\quad \ n \rightarrow \infty,\ T \rightarrow 0 \quad} 
	\rho_{x_i,s_j} \\
	\hat{\rho}_{n,T}(\Delta^{R}_{i,\tau},\Delta^{v}_{j}) \xrightarrow[p]{\quad \ n \rightarrow \infty, \ T \rightarrow 0 \quad} 
	\rho_{x_i,v_j}
\end{gather}

\subsection{Dupire's Local Volatility Model}

The linear system for G2++/Dupire \parencite{37} is 2x2, as local volatility model has only one state variable. Theorem \ref{thm2} still holds but only equation (\ref{bs_only}) is required.

For G1++/Dupire, the empirical correlation converges to the instantaneous correlation.
\begin{gather}
	\hat{\rho}_{n,T}(\Delta^{R}_{i,\tau},\Delta^{s}_{j}) \xrightarrow[p]{\quad \ n \rightarrow \infty,\ T \rightarrow 0 \quad} 
	\rho_{x_i,s_j}
\end{gather}

\subsection{Jump Diffusion Process}

Let $X_t,Y_t$ be jump diffusion processes, with the continuous parts denoted by $X_t^c, Y_t^c$ respectively. The quadratic covariation $[X,Y]_T$ can be expressed in terms of the quadratic covariation between the continuous parts and jumps.

\begin{equation}
	[X,Y]_T = [X^c,Y^c]_T + \sum_{t \le T} \Delta X_t \Delta Y_t
\end{equation}
where $\Delta X_t, \Delta Y_t $ denotes the jumps of $X,Y$ respectively at time $t$. The dynamics of \textcite{18} is given by
\begin{gather}
	ds_t = \left(\tilde{r}_t-\tilde{q}_t - \lambda\left(\exp(\mu+\sigma^2/2)-1\right)-v_t/2\right)dt + \sqrt{v_t} dW_t^s + \log(J_t)dN_t \\
	dv_t = \kappa(\theta - v_t) dt + \xi \sqrt{v_t} dW_t^v \\
	dW_t^s dW_t^v = \rho_{s,v} \ dt\\
	N_t \sim \mbox{independent Poisson}(\lambda)\\
	\log(J_t) \sim N(\mu, \sigma)
\end{gather}

For the case of G2++/Bates(\citeyear{18}), as the spot rate $R^\tau(t)$ is continuous, $\sum_{t \le T} \Delta R^\tau_i(t) \Delta s_j(t) = 0$ almost surely. When both models are Bates, since the two jump processes are assumed to be independent, they have no common jump term (no co-jump); hence $\sum_{t \le T} \Delta s_i(t) \Delta s_j(t) = 0$ also. Therefore, the results in theorem \ref{thm2} and \ref{thm3} still hold.

\subsection{Stochastic Interest Rate}
For an equity/FX models with a stochastic short rate, the hybrid system is coupled as follows:
\begin{gather}
	r_t = \phi(t) + x_t + y_t \\
	dx_t = -a x_t dt + \sigma dW_t^x, \ x_0 = 0 \\
	dy_t = -b y_t dt + \eta dW_t^y, \ y_0 = 0 \\
	dW_t^xdW_t^y = \rho_{x,y} \\
	ds_t = (r_t-\tilde{q}_t-v_t/2)dt + \sqrt{v_t} dW_t^s \\
	dv_t = \kappa(\theta - v_t) dt + \xi \sqrt{v_t} dW_t^v  \\
	dW_t^s dW_t^v = \rho_{s,v} 
\end{gather}

Since stochastic interest rate only affects the finite variation $dt$ term in $s_t$, the quadratic co-variation is unchanged. Hence, the results in theorem \ref{thm2} and \ref{thm3} are not affected. For the case of FX Heston model, $\tilde{q}_t$ can also be stochastic, but again it has no effect on previous results. 

\section{Numerical Study} \label{sec:num}

In this section, the accuracy and robustness of the method is established by comparing the estimated correlations with known values. 1,000 sample paths of a given hybrid model will be simulated using the parameters in Table \ref{param}. Each sample path consists of $n$ time steps with size $dt$. In each sample path, the instantaneous correlation is estimated using theorem \ref{thm1}, \ref{thm2}, and \ref{thm3}. With those 1,000 sample correlation estimates, the bias = (mean(estimates) - known value) and standard error = stdev(estimates) are computed.

\begin{table}[!ht]
	\centering
	\caption{Model Parameters}
	\label{param}
	\begin{tabular}{ccc}
		\toprule
		G2++/G2++ & G2++/Heston & Heston/Heston \\
		\midrule
		$a$=\{0.1,0.15\}, $b$=\{0.2,0.25\} & $a$=0.1, $b$=0.2 & $\kappa$=\{1,1.1\}, $\theta$=\{0.2,0.22\}\\
		$\sigma$=\{0.01,0.015\}, $\eta$=\{0.02,0.025\} & $\sigma$=0.01, $\eta$=0.02 & $\xi$=\{0.3,0.33\}, $v_0$=\{0.1,0.11\}\\
		$\rho$=\{0.5,0.55\} & $\rho_{G2}$=0.5 & $\rho$=\{-0.8,-0.88\}\\
								& $\kappa$=1, $\theta$=0.2 & \\
								& $\xi$=0.3, $v_0$=0.1, $\rho_{Heston}$=-0.8 & \\
		$\Sigma=\begin{bmatrix} 0.1 & 0.2 \\ 0.3 & 0.4 \end{bmatrix}$ & $\Sigma=\begin{bmatrix} 0.1 & -0.2 \\ 0.3 & -0.4 \end{bmatrix}$ & $\Sigma=\begin{bmatrix} 0.1 & -0.2 \\ -0.3 & 0.4 \end{bmatrix}$\\
		\bottomrule
	\end{tabular}
\end{table}

For the cases of G2++/Heston and Heston/Heston, the sampling period $T$ is required to be short while number of time steps $n$ to be large. Hence, it will be ideal to use intraday data for the estimation. However, intraday data may not be readily available, especially for implied volatility, so statistics using both $dt = 0.01/250$ (4-minute snapshot assuming 6.5 hours trading day) and $dt = 1/250$ (daily) are computed for the the sake of comparison, with $n$ ranging from 100 to 10,000. For 10,000 time steps, daily data corresponds to 40 years of history, while that of intraday data is only 5 months.

\subsection{Known Parameters}

In order to infer the instantaneous correlations in theorem \ref{thm1} and \ref{thm2}, the parameters of G2++, which are typically calibrated from the observed volatility cube of swaptions, are required. Nonetheless, in this section, those parameters are assumed to be known exactly, and the effect of misspecified model parameters will be examined in the next section.

\begin{table}[!ht]
	\centering
	\caption{Estimation Statistics with Known Parameters: G2++/G2++ \\(1,000 samples, top: bias, bottom: standard error)}
	\label{g2g2}	
	\begin{tabular}{c!{\vrule width 1pt}cccc!{\vrule width 1pt}cccc}
		\toprule
		&  \multicolumn{4}{c!{\vrule width 1pt}}{ step size dt = 0.01 day} & \multicolumn{4}{c}{ step size dt = 1 day} \\[5pt]
		\# time steps n & 
		\parbox{0.5in}{\centering $\hat{\rho}_{x0,x1}$ \\ 10\%} & \parbox{0.5in}{\centering $\hat{\rho}_{x0,y1}$ \\ 20\%} & 
		\parbox{0.5in}{\centering $\hat{\rho}_{y0,x1}$ \\ 30\%} & \parbox{0.5in}{\centering $\hat{\rho}_{y0,y1}$ \\ 40\%} & 
		\parbox{0.5in}{\centering $\hat{\rho}_{x0,x1}$ \\ 10\%} & \parbox{0.5in}{\centering $\hat{\rho}_{x0,y1}$ \\ 20\%} & 
		\parbox{0.5in}{\centering $\hat{\rho}_{y0,x1}$ \\ 30\%} & \parbox{0.5in}{\centering $\hat{\rho}_{y0,y1}$ \\ 40\%} \\[10pt]
		\midrule
		100&\parbox{0.5in}{\centering -0.08\%\\(9.83\%)} & \parbox{0.5in}{\centering 0.36\%\\(9.63\%)} & \parbox{0.5in}{\centering -0.30\%\\(8.84\%)} & \parbox{0.5in}{\centering 0.21\%\\(8.24\%)}&\parbox{0.5in}{\centering -0.08\%\\(9.83\%)} & \parbox{0.5in}{\centering 0.36\%\\(9.63\%)} & \parbox{0.5in}{\centering -0.30\%\\(8.84\%)} & \parbox{0.5in}{\centering 0.21\%\\(8.24\%)}\\[15pt]
		1,000&\parbox{0.5in}{\centering -0.07\%\\(3.15\%)} & \parbox{0.5in}{\centering -0.15\%\\(2.92\%)} & \parbox{0.5in}{\centering 0.12\%\\(2.79\%)} & \parbox{0.5in}{\centering -0.04\%\\(2.64\%)}&\parbox{0.5in}{\centering -0.07\%\\(3.15\%)} & \parbox{0.5in}{\centering -0.15\%\\(2.92\%)} & \parbox{0.5in}{\centering 0.12\%\\(2.79\%)} & \parbox{0.5in}{\centering -0.04\%\\(2.64\%)}\\[15pt]
		2,500&\parbox{0.5in}{\centering -0.01\%\\(2.03\%)} & \parbox{0.5in}{\centering -0.08\%\\(1.99\%)} & \parbox{0.5in}{\centering -0.06\%\\(1.75\%)} & \parbox{0.5in}{\centering -0.10\%\\(1.72\%)}&\parbox{0.5in}{\centering -0.01\%\\(2.03\%)} & \parbox{0.5in}{\centering -0.08\%\\(1.99\%)} & \parbox{0.5in}{\centering -0.06\%\\(1.75\%)} & \parbox{0.5in}{\centering -0.10\%\\(1.72\%)}\\[15pt]
		5,000&\parbox{0.5in}{\centering -0.10\%\\(1.41\%)} & \parbox{0.5in}{\centering -0.02\%\\(1.30\%)} & \parbox{0.5in}{\centering 0.00\%\\(1.30\%)} & \parbox{0.5in}{\centering 0.06\%\\(1.14\%)}&\parbox{0.5in}{\centering -0.10\%\\(1.41\%)} & \parbox{0.5in}{\centering -0.02\%\\(1.30\%)} & \parbox{0.5in}{\centering -0.01\%\\(1.30\%)} & \parbox{0.5in}{\centering 0.06\%\\(1.14\%)}\\[15pt]
		10,000&\parbox{0.5in}{\centering 0.02\%\\(0.95\%)} & \parbox{0.5in}{\centering 0.04\%\\(0.95\%)} & \parbox{0.5in}{\centering 0.04\%\\(0.88\%)} & \parbox{0.5in}{\centering 0.07\%\\(0.79\%)}&\parbox{0.5in}{\centering 0.02\%\\(0.95\%)} & \parbox{0.5in}{\centering 0.04\%\\(0.95\%)} & \parbox{0.5in}{\centering 0.04\%\\(0.88\%)} & \parbox{0.5in}{\centering 0.07\%\\(0.79\%)}\\[15pt]
		\bottomrule
	\end{tabular}
\end{table}

By theorem \ref{thm1}, the step size $dt$ has no effect on the estimation, so the number on the left and right panels in Table \ref{g2g2} are exactly the same. The biases are essentially 0, and those small numbers are simply statistical error due to finite sample size. On the other hand, the standard errors reduce significantly from around 10\% to 1\% when the number of time steps $n$ increases from 100 to 10,000. The standard error seems not sensitive to the underlying correlation value, as they are in the same ballpark whether the true correlation is 10\% or 40\%, given the same $n$.

\begin{table}[!ht]
	\centering
	\caption{Estimation Statistics with Known Parameters: G2++/Heston \\(1,000 samples, top: bias, bottom: standard error)}
	\label{g2heston}	
	\begin{tabular}{c!{\vrule width 1pt}cccc!{\vrule width 1pt}cccc}
		\toprule
		&  \multicolumn{4}{c!{\vrule width 1pt}}{ step size dt = 0.01 day} & \multicolumn{4}{c}{ step size dt = 1 day} \\[5pt]
		\# time steps n & 
		\parbox{0.5in}{\centering $\hat{\rho}_{x0,s1}$ \\ 10\%} & \parbox{0.5in}{\centering $\hat{\rho}_{y0,s1}$ \\ 30\%} & 
		\parbox{0.5in}{\centering $\hat{\rho}_{x0,v1}$ \\ -20\%} & \parbox{0.5in}{\centering $\hat{\rho}_{y0,v1}$ \\ -40\%} & 
		\parbox{0.5in}{\centering $\hat{\rho}_{x0,s1}$ \\ 10\%} & \parbox{0.5in}{\centering $\hat{\rho}_{y0,s1}$ \\ 30\%} & 
		\parbox{0.5in}{\centering $\hat{\rho}_{x0,v1}$ \\ -20\%} & \parbox{0.5in}{\centering $\hat{\rho}_{y0,v1}$ \\ -40\%} \\[10pt]
		\midrule
		100&\parbox{0.5in}{\centering -0.06\%\\(9.82\%)} & \parbox{0.5in}{\centering -0.31\%\\(9.10\%)} & \parbox{0.5in}{\centering 0.42\%\\(9.78\%)} & \parbox{0.5in}{\centering 0.56\%\\(8.62\%)}&\parbox{0.5in}{\centering -0.08\%\\(9.79\%)} & \parbox{0.5in}{\centering -0.44\%\\(9.12\%)} & \parbox{0.5in}{\centering 0.51\%\\(9.78\%)} & \parbox{0.5in}{\centering 0.77\%\\(8.60\%)}\\[15pt]
		1,000&\parbox{0.5in}{\centering -0.07\%\\(3.15\%)} & \parbox{0.5in}{\centering 0.11\%\\(2.83\%)} & \parbox{0.5in}{\centering -0.04\%\\(3.07\%)} & \parbox{0.5in}{\centering -0.14\%\\(2.63\%)}&\parbox{0.5in}{\centering -0.29\%\\(3.16\%)} & \parbox{0.5in}{\centering -0.49\%\\(2.85\%)} & \parbox{0.5in}{\centering 0.37\%\\(3.09\%)} & \parbox{0.5in}{\centering 0.65\%\\(2.67\%)}\\[15pt]
		2,500&\parbox{0.5in}{\centering -0.03\%\\(2.03\%)} & \parbox{0.5in}{\centering -0.12\%\\(1.77\%)} & \parbox{0.5in}{\centering 0.00\%\\(1.89\%)} & \parbox{0.5in}{\centering 0.08\%\\(1.67\%)}&\parbox{0.5in}{\centering -0.24\%\\(2.03\%)} & \parbox{0.5in}{\centering -0.78\%\\(1.79\%)} & \parbox{0.5in}{\centering 0.45\%\\(1.91\%)} & \parbox{0.5in}{\centering 0.98\%\\(1.70\%)}\\[15pt]
		5,000&\parbox{0.5in}{\centering -0.13\%\\(1.42\%)} & \parbox{0.5in}{\centering -0.11\%\\(1.33\%)} & \parbox{0.5in}{\centering 0.17\%\\(1.37\%)} & \parbox{0.5in}{\centering 0.18\%\\(1.26\%)}&\parbox{0.5in}{\centering -0.36\%\\(1.42\%)} & \parbox{0.5in}{\centering -0.79\%\\(1.35\%)} & \parbox{0.5in}{\centering 0.65\%\\(1.39\%)} & \parbox{0.5in}{\centering 1.11\%\\(1.30\%)}\\[15pt]
		10,000&\parbox{0.5in}{\centering -0.04\%\\(0.96\%)} & \parbox{0.5in}{\centering -0.14\%\\(0.91\%)} & \parbox{0.5in}{\centering 0.13\%\\(0.94\%)} & \parbox{0.5in}{\centering 0.25\%\\(0.87\%)}&\parbox{0.5in}{\centering -0.25\%\\(0.97\%)} & \parbox{0.5in}{\centering -0.77\%\\(0.92\%)} & \parbox{0.5in}{\centering 0.55\%\\(0.94\%)} & \parbox{0.5in}{\centering 1.10\%\\(0.88\%)}\\[15pt]
		\bottomrule
	\end{tabular}
\end{table}

In the case of G2++/Heston, $dt$ matters, so one can see the biases on daily data (right) are higher than that of intraday data (left) in Table \ref{g2heston}. However, the difference is not substantial (< 1\%). In contrast, the standard error is not affected by $dt$, and again driven mostly by $n$. 

Therefore, it is recommended to use a longer time series to reduce the standard error since the bias due to large $T$ is negligible. To obtain a reasonable estimate, one should use at least 1,000 time steps, which amounts to 4 years of daily data and 2 weeks of intraday data. 

Correlation is in general not constant. With the use of intraday data, one can observe the trend in correlation by using samples over a very short period of time. However, this is not possible with daily data. For example, 2 weeks of daily data will likely produce some garbage result, while 2 weeks of intraday data may give a quite decent estimate.

\begin{table}[!ht]
	\centering
	\caption{Estimation Statistics with Known Parameters: Heston/Heston \\(1,000 samples, top: bias, bottom: standard error)}
	\label{hestonheston}	
	\begin{tabular}{c!{\vrule width 1pt}cccc!{\vrule width 1pt}cccc}
		\toprule
		&  \multicolumn{4}{c!{\vrule width 1pt}}{ step size dt = 0.01 day} & \multicolumn{4}{c}{ step size dt = 1 day} \\[5pt]
		\# time steps n & 
		\parbox{0.5in}{\centering $\hat{\rho}_{s0,s1}$ \\ 10\%} & \parbox{0.5in}{\centering $\hat{\rho}_{v0,s1}$ \\ -30\%} & 
		\parbox{0.5in}{\centering $\hat{\rho}_{s0,v1}$ \\ -20\%} & \parbox{0.5in}{\centering $\hat{\rho}_{v0,v1}$ \\ 40\%} & 
		\parbox{0.5in}{\centering $\hat{\rho}_{s0,s1}$ \\ 10\%} & \parbox{0.5in}{\centering $\hat{\rho}_{v0,s1}$ \\ -30\%} & 
		\parbox{0.5in}{\centering $\hat{\rho}_{s0,v1}$ \\ -20\%} & \parbox{0.5in}{\centering $\hat{\rho}_{v0,v1}$ \\ 40\%} \\[10pt]
		\midrule
		100&\parbox{0.5in}{\centering -0.08\%\\(9.60\%)} & \parbox{0.5in}{\centering 0.01\%\\(9.00\%)} & \parbox{0.5in}{\centering 0.44\%\\(9.73\%)} & \parbox{0.5in}{\centering -0.30\%\\(8.64\%)}&\parbox{0.5in}{\centering -0.14\%\\(9.68\%)} & \parbox{0.5in}{\centering 0.16\%\\(9.07\%)} & \parbox{0.5in}{\centering 0.58\%\\(9.81\%)} & \parbox{0.5in}{\centering -0.52\%\\(8.72\%)}\\[15pt]
		1,000&\parbox{0.5in}{\centering -0.08\%\\(3.18\%)} & \parbox{0.5in}{\centering 0.17\%\\(2.91\%)} & \parbox{0.5in}{\centering -0.01\%\\(3.11\%)} & \parbox{0.5in}{\centering -0.10\%\\(2.74\%)}&\parbox{0.5in}{\centering -0.33\%\\(3.25\%)} & \parbox{0.5in}{\centering 0.88\%\\(3.04\%)} & \parbox{0.5in}{\centering 0.45\%\\(3.21\%)} & \parbox{0.5in}{\centering -0.98\%\\(2.90\%)}\\[15pt]
		2,500&\parbox{0.5in}{\centering -0.02\%\\(2.00\%)} & \parbox{0.5in}{\centering 0.04\%\\(1.82\%)} & \parbox{0.5in}{\centering 0.00\%\\(1.91\%)} & \parbox{0.5in}{\centering -0.04\%\\(1.67\%)}&\parbox{0.5in}{\centering -0.26\%\\(2.07\%)} & \parbox{0.5in}{\centering 0.83\%\\(1.92\%)} & \parbox{0.5in}{\centering 0.53\%\\(1.99\%)} & \parbox{0.5in}{\centering -1.04\%\\(1.79\%)}\\[15pt]
		5,000&\parbox{0.5in}{\centering -0.13\%\\(1.44\%)} & \parbox{0.5in}{\centering 0.22\%\\(1.30\%)} & \parbox{0.5in}{\centering 0.16\%\\(1.42\%)} & \parbox{0.5in}{\centering -0.24\%\\(1.21\%)}&\parbox{0.5in}{\centering -0.38\%\\(1.51\%)} & \parbox{0.5in}{\centering 1.04\%\\(1.39\%)} & \parbox{0.5in}{\centering 0.71\%\\(1.48\%)} & \parbox{0.5in}{\centering -1.29\%\\(1.31\%)}\\[15pt]
		10,000&\parbox{0.5in}{\centering -0.03\%\\(0.97\%)} & \parbox{0.5in}{\centering 0.20\%\\(0.91\%)} & \parbox{0.5in}{\centering 0.13\%\\(0.96\%)} & \parbox{0.5in}{\centering -0.28\%\\(0.88\%)}&\parbox{0.5in}{\centering -0.26\%\\(1.00\%)} & \parbox{0.5in}{\centering 0.98\%\\(0.96\%)} & \parbox{0.5in}{\centering 0.63\%\\(0.99\%)} & \parbox{0.5in}{\centering -1.26\%\\(0.93\%)}\\[15pt]
		\bottomrule
	\end{tabular}
\end{table}

The result of Heston/Heston in Table \ref{hestonheston} is similar to that of G2++/Heston: $dt$ matters but not very significantly, while larger $n$ can drastically reduce the standard error. Again, the standard error does not depend on the sign and magnitude of the true correlation.

\subsection{Misspecified Parameters}

In practice, the G2++ parameters cannot be calibrated with 100\% accuracy as there will be error in discretization and optimization (not enough iterations, local minimum instead of global minimum etc.). In addition, some values in the volatility cube may be stale due to inactive trading. As a result, the correlations estimated from theorem \ref{thm1} and \ref{thm2} may be compromised, as one is using the wrong parameters in the simultaneous equations.

In this section, the instantaneous correlation will be inferred from wrong model parameters that are under-estimated by 30\%.

\begin{table}[!ht]
	\centering
	\caption{Estimation Statistics with 30\% Under-Estimated Parameters: G2++/G2++ \\(1,000 samples, top: bias, bottom: standard error)}
	\label{g2g2_}	
	\begin{tabular}{c!{\vrule width 1pt}cccc!{\vrule width 1pt}cccc}
		\toprule
		&  \multicolumn{4}{c!{\vrule width 1pt}}{ step size dt = 0.01 day} & \multicolumn{4}{c}{ step size dt = 1 day} \\[5pt]
		\# time steps n & 
		\parbox{0.5in}{\centering $\hat{\rho}_{x0,x1}$ \\ 10\%} & \parbox{0.5in}{\centering $\hat{\rho}_{x0,y1}$ \\ 20\%} & 
		\parbox{0.5in}{\centering $\hat{\rho}_{y0,x1}$ \\ 30\%} & \parbox{0.5in}{\centering $\hat{\rho}_{y0,y1}$ \\ 40\%} & 
		\parbox{0.5in}{\centering $\hat{\rho}_{x0,x1}$ \\ 10\%} & \parbox{0.5in}{\centering $\hat{\rho}_{x0,y1}$ \\ 20\%} & 
		\parbox{0.5in}{\centering $\hat{\rho}_{y0,x1}$ \\ 30\%} & \parbox{0.5in}{\centering $\hat{\rho}_{y0,y1}$ \\ 40\%} \\[10pt]
		\midrule
		100&\parbox{0.5in}{\centering -0.29\%\\(8.30\%)} & \parbox{0.5in}{\centering -2.24\%\\(8.31\%)} & \parbox{0.5in}{\centering -2.59\%\\(7.62\%)} & \parbox{0.5in}{\centering -3.58\%\\(7.29\%)}&\parbox{0.5in}{\centering -0.29\%\\(8.30\%)} & \parbox{0.5in}{\centering -2.24\%\\(8.31\%)} & \parbox{0.5in}{\centering -2.59\%\\(7.62\%)} & \parbox{0.5in}{\centering -3.58\%\\(7.29\%)}\\[15pt]
		1,000&\parbox{0.5in}{\centering -0.31\%\\(2.65\%)} & \parbox{0.5in}{\centering -2.65\%\\(2.54\%)} & \parbox{0.5in}{\centering -2.28\%\\(2.42\%)} & \parbox{0.5in}{\centering -3.77\%\\(2.35\%)}&\parbox{0.5in}{\centering -0.31\%\\(2.65\%)} & \parbox{0.5in}{\centering -2.65\%\\(2.54\%)} & \parbox{0.5in}{\centering -2.28\%\\(2.42\%)} & \parbox{0.5in}{\centering -3.77\%\\(2.35\%)}\\[15pt]
		2,500&\parbox{0.5in}{\centering -0.26\%\\(1.72\%)} & \parbox{0.5in}{\centering -2.59\%\\(1.73\%)} & \parbox{0.5in}{\centering -2.43\%\\(1.53\%)} & \parbox{0.5in}{\centering -3.83\%\\(1.54\%)}&\parbox{0.5in}{\centering -0.26\%\\(1.73\%)} & \parbox{0.5in}{\centering -2.59\%\\(1.73\%)} & \parbox{0.5in}{\centering -2.43\%\\(1.53\%)} & \parbox{0.5in}{\centering -3.83\%\\(1.54\%)}\\[15pt]
		5,000&\parbox{0.5in}{\centering -0.33\%\\(1.19\%)} & \parbox{0.5in}{\centering -2.55\%\\(1.13\%)} & \parbox{0.5in}{\centering -2.37\%\\(1.12\%)} & \parbox{0.5in}{\centering -3.69\%\\(1.01\%)}&\parbox{0.5in}{\centering -0.33\%\\(1.19\%)} & \parbox{0.5in}{\centering -2.55\%\\(1.13\%)} & \parbox{0.5in}{\centering -2.37\%\\(1.12\%)} & \parbox{0.5in}{\centering -3.69\%\\(1.01\%)}\\[15pt]
		10,000&\parbox{0.5in}{\centering -0.23\%\\(0.81\%)} & \parbox{0.5in}{\centering -2.49\%\\(0.82\%)} & \parbox{0.5in}{\centering -2.33\%\\(0.76\%)} & \parbox{0.5in}{\centering -3.68\%\\(0.71\%)}&\parbox{0.5in}{\centering -0.23\%\\(0.81\%)} & \parbox{0.5in}{\centering -2.49\%\\(0.82\%)} & \parbox{0.5in}{\centering -2.33\%\\(0.76\%)} & \parbox{0.5in}{\centering -3.68\%\\(0.71\%)}\\[15pt]
		\bottomrule
	\end{tabular}
\end{table}

\begin{table}[!ht]
	\centering
	\caption{Estimation Statistics with 30\% Under-Estimated Parameters: G2++/Heston \\(1,000 samples, top: bias, bottom: standard error)}
	\label{g2heston_}	
	\begin{tabular}{c!{\vrule width 1pt}cccc!{\vrule width 1pt}cccc}
		\toprule
		&  \multicolumn{4}{c!{\vrule width 1pt}}{ step size dt = 0.01 day} & \multicolumn{4}{c}{ step size dt = 1 day} \\[5pt]
		\# time steps n & 
		\parbox{0.5in}{\centering $\hat{\rho}_{x0,s1}$ \\ 10\%} & \parbox{0.5in}{\centering $\hat{\rho}_{y0,s1}$ \\ 30\%} & 
		\parbox{0.5in}{\centering $\hat{\rho}_{x0,v1}$ \\ -20\%} & \parbox{0.5in}{\centering $\hat{\rho}_{y0,v1}$ \\ -40\%} & 
		\parbox{0.5in}{\centering $\hat{\rho}_{x0,s1}$ \\ 10\%} & \parbox{0.5in}{\centering $\hat{\rho}_{y0,s1}$ \\ 30\%} & 
		\parbox{0.5in}{\centering $\hat{\rho}_{x0,v1}$ \\ -20\%} & \parbox{0.5in}{\centering $\hat{\rho}_{y0,v1}$ \\ -40\%} \\[10pt]
		\midrule
		100&\parbox{0.5in}{\centering -0.50\%\\(9.14\%)} & \parbox{0.5in}{\centering -1.83\%\\(8.72\%)} & \parbox{0.5in}{\centering 1.47\%\\(9.11\%)} & \parbox{0.5in}{\centering 2.25\%\\(8.26\%)}&\parbox{0.5in}{\centering -0.52\%\\(9.12\%)} & \parbox{0.5in}{\centering -1.95\%\\(8.74\%)} & \parbox{0.5in}{\centering 1.56\%\\(9.11\%)} & \parbox{0.5in}{\centering 2.45\%\\(8.24\%)}\\[15pt]
		1,000&\parbox{0.5in}{\centering -0.51\%\\(2.94\%)} & \parbox{0.5in}{\centering -1.44\%\\(2.72\%)} & \parbox{0.5in}{\centering 1.04\%\\(2.86\%)} & \parbox{0.5in}{\centering 1.57\%\\(2.52\%)}&\parbox{0.5in}{\centering -0.72\%\\(2.95\%)} & \parbox{0.5in}{\centering -2.01\%\\(2.75\%)} & \parbox{0.5in}{\centering 1.43\%\\(2.87\%)} & \parbox{0.5in}{\centering 2.33\%\\(2.57\%)}\\[15pt]
		2,500&\parbox{0.5in}{\centering -0.47\%\\(1.89\%)} & \parbox{0.5in}{\centering -1.65\%\\(1.71\%)} & \parbox{0.5in}{\centering 1.08\%\\(1.76\%)} & \parbox{0.5in}{\centering 1.78\%\\(1.60\%)}&\parbox{0.5in}{\centering -0.67\%\\(1.89\%)} & \parbox{0.5in}{\centering -2.28\%\\(1.72\%)} & \parbox{0.5in}{\centering 1.50\%\\(1.78\%)} & \parbox{0.5in}{\centering 2.64\%\\(1.63\%)}\\[15pt]
		5,000&\parbox{0.5in}{\centering -0.57\%\\(1.32\%)} & \parbox{0.5in}{\centering -1.65\%\\(1.28\%)} & \parbox{0.5in}{\centering 1.23\%\\(1.28\%)} & \parbox{0.5in}{\centering 1.88\%\\(1.21\%)}&\parbox{0.5in}{\centering -0.79\%\\(1.32\%)} & \parbox{0.5in}{\centering -2.29\%\\(1.30\%)} & \parbox{0.5in}{\centering 1.69\%\\(1.29\%)} & \parbox{0.5in}{\centering 2.77\%\\(1.24\%)}\\[15pt]
		10,000&\parbox{0.5in}{\centering -0.48\%\\(0.90\%)} & \parbox{0.5in}{\centering -1.67\%\\(0.87\%)} & \parbox{0.5in}{\centering 1.20\%\\(0.87\%)} & \parbox{0.5in}{\centering 1.94\%\\(0.83\%)}&\parbox{0.5in}{\centering -0.68\%\\(0.90\%)} & \parbox{0.5in}{\centering -2.27\%\\(0.88\%)} & \parbox{0.5in}{\centering 1.60\%\\(0.88\%)} & \parbox{0.5in}{\centering 2.76\%\\(0.84\%)}\\[15pt]
		\bottomrule
	\end{tabular}
\end{table}

As shown in Table \ref{g2g2_}, the biases of G2++/G2++ go from essentially 0 to the maximum of around 4\%, and they also depend on the underlying correlation values. For the case of the 10\% correlation, the bias is only 0.3\% while that of the 40\% correlation is around 4\%. Nonetheless, the biases are still quite manageable, given all parameters are significantly under-estimated by 30\%. The standard error is very close to the case of known parameters, and again driven mostly by $n$.

The effect of misspecified parameters in G2++/Heston is in general smaller than that of G2++/G2++ as the parameters in Heston are not required to infer the instantaneous correlations.

\section{Conclusion} \label{sec:con}
In this article, a simple method is introduced to estimate correlations in a hybrid system, consisting of affine term-structure, stochastic volatility, local volatility and jump-diffusion models, with or without stochastic interest rate. The method is extremely fast as no high-dimensional optimization with equality constraints is involved; one only needs to solve some low-dimension linear systems and a bisection-style algorithm in shrinking.

Theoretically, the sampling period $T$ is required to be short under G2++/Heston and Heston/Heston, but numerical study shows that the bias introduced by finite $T$ is insignificant. The crucial factor in determining the accuracy is indeed the time series sample size $n$. For a reasonable estimate, it is recommended to use at least 1,000 data points, which amount to 2 weeks of intraday data (4-minute snapshot) or 4 years of daily data. This method is also robust to the calibration errors of the interest rate model parameters.

%\bibliographystyle{unsrtnat2}
%\bibliographystyle{apacite}
%\bibliography{output}
\printbibliography

\end{document}